\documentclass[aip,jap,reprint]{revtex4-1}
\usepackage{graphicx}% Include figure files
\usepackage{dcolumn}% Align table columns on decimal point
\usepackage{bm}% bold math

\begin{document}

\title{Strain-induced magnetism in MoS$_2$ monolayer with defects}

\author{Peng Tao}
\author{Huaihong Guo}
\author{Teng Yang}
\email[E-mail: ]{yangteng@imr.ac.cn}
\author{Zhidong Zhang}
\affiliation{Shenyang National Laboratory for Materials Science,
Institute of Metal Research and International Centre for Materials Physics,
Chinese Academy of Sciences, 72 Wenhua Road, Shenyang 110016,
PRC}

\date{\today}

%---------------------------------------------------------------------
\begin{abstract}
The strain-induced magnetism is observed in single-layer MoS$_2$ with atomic single vacancies from density functional calculations. Calculated magnetic moment is no less than 2 $\mu_B$ per vacancy defect. The strain-induced band gap closure is concurrent with the occurrence of the magnetism. Possible physical mechanism of the emergence of strain-induced magnetism is illustrated. We also demonstrate the possibility to test the predicted magnetism in experiment. Our study may provide an opportunity for the design of new type of memory-switching or logic devices by using earth-rich nonmagnetic materials MoS$_2$.
\end{abstract}
%---------------------------------------------------------------------
%\pacs{
%61.72.J-,  % Point defects
%61.72.Qq,  % Voids (crystal defects)
%71.15.Mb,  % Self-consistent field calculations for solids
%73.20.Hb,  % Adsorbates electron states
%75.70.Rf  % Surface magnetism
%}

\maketitle
%---------------------------------------------------------------------
\section{introduction}
Vacancy-induced magnetism in graphene \cite{graphenemag,rpp73.056501,PhysRevFocus.25.6,naturephysics.8.199} has attracted tremendous attention lately as a completely new generation of materials for spintronics. The possible physical mechanisms have also been intensively studied\cite{PhysRevB.75.125408,rpp73.056501,PhysRevB.77.195428, PhysRevLett.93.187202,JPSJ.76.064713}. Analogous to graphene, single-layer MoS$_2$ has been demonstrated to be powerful in nanoelectronic applications\cite{mobility,on-off-ratio,SMT,PhysRevB.85.033305,indirect-band-gap-experiment}. Meanwhile, atomic vacancy defects in single-layer MoS$_2$\cite{Coleman04022011,prepare} can be easily and even controllably obtained by chemical vapor deposition(CVD)\cite{cvd}, field evaporation\cite{stm1,stm2} and electron irradiation method\cite{PhysRevLett.109.035503}. However, reports have barely been found on magnetism appearing in MoS$_2$ with single vacancy\cite{C3CP50381J,jp2000442}.

In this work, we studied a possible emergence of magnetism in MoS$_2$ monolayer with S$_2$ or Mo single vacancy under strain. The idea of applying strain arises from the following considerations: first, the magnetism found in the defected graphene may arise partly from an itinerant origin \cite{PhysRevB.75.125408, rpp73.056501, santos} in which conducting electrons play a role. In this regard, the absence of magnetism in defected MoS$_2$ is possibly due to the existence of a sizable band gap; secondly, tensile strain has been proven to be efficient in reducing band gap of pristine single-layer MoS$_2$ (P-MoS$_2$) \cite{SMT,indirect-band-gap-experiment} and in improving carrier mobility as well \cite{mobility,PhysRevB.85.033305}. From our calculation, indeed, tensile strain can not only help close the band gap of the defected MoS$_2$, but also give rise to a vacancy-related magnetism. More importantly, we found a concurrent appearance of the strain-induced band gap closure and the magnetism. Our analysis shows that the emergence of magnetic moments does have some relevance to the delocalized electronic properties in those defected structures.

\section{computational method}
We performed first-principles plane wave calculations within density functional theory(DFT) using the projector augmented wave(PAW)\cite{paw} pseudopotential as implemented within the Vienna Ab-initio Simulation Package(VASP)\cite{vasp}. The electronic exchange-correlation potential was treated by the generalized gradient approximation(GGA) with the Perdew-Burke-Ernzerhof(PBE) flavor \cite{pbe}. We used the DFT-D2 method of Grimme\cite{vdw} to describe Van der Waals(VdW) interactions between vacancies. A vacuum spacing of 20 \AA\ was used to avoid interaction between MoS$_2$ layers. Electronic kinetic energy cutoff was set to be 450 eV. Self-consistency was reached once energy difference between two consecutive steps was smaller than $10^{-5}$ eV, and atomic Hellman-Feynman force was minimized less than 0.01 eV/\AA. In the self-consistent total energy calculations the Brillouin zone was sampled by 7$\times$7$\times$1 k mesh. We employed the fixed spin moment method \cite{fsm1,fsm2} to search for the most stable spin configuration and meanwhile to calculate the corresponding value of spin moment. Biaxial strain rather than uniaxial strain was applied due to a practical consideration that strain arising from the coupling of monolayer MoS$_2$ to substrate is usually isotropic. Our calculated lattice constant of P-MoS$_2$ was 3.19 \AA, which is slightly larger than experimental value of 3.16 \AA. A direct band gap of 1.67 eV, a little smaller than experimental value\cite{PhysRevLett.105.136805}, was obtained. Both values are reasonably acceptable since the GGA approximation usually underestimates the band gap size and the binding strength. Several supercell sizes (3$\times$3, 4$\times$4 and etc.) were tested but vacancy stability remains unchanged as supercell size goes bigger than 4$\times$4. So 4$\times$4 was chosen all through this manuscript.

%--------------------------------------------------------------------
\begin{figure}[t]
\includegraphics[width=0.8\columnwidth]{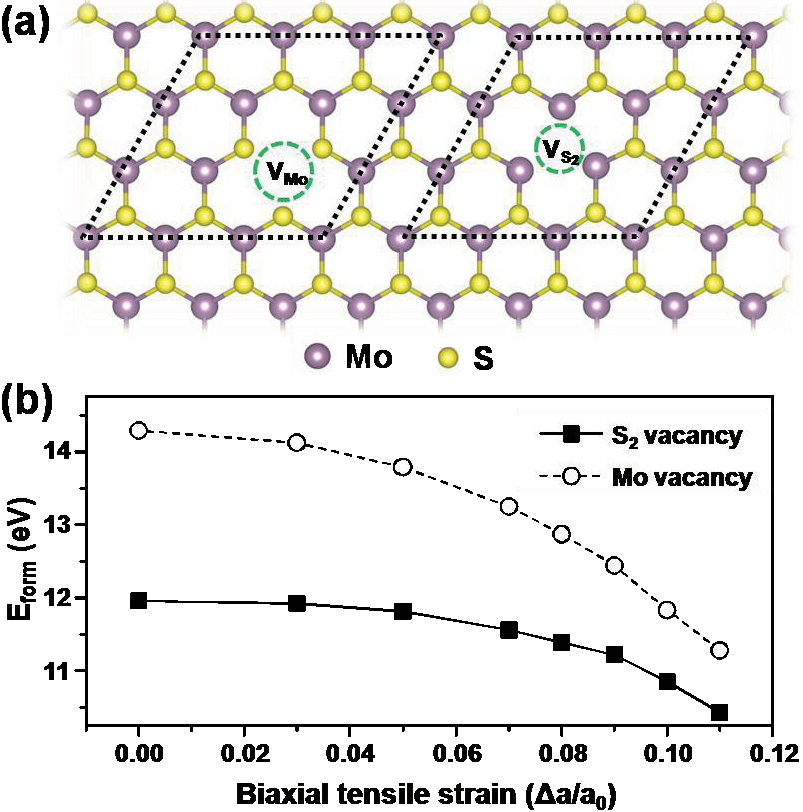}
\caption{Schematics of single-layer MoS$_2$ with atomic vacancy. (a) The Mo or S$_2$ vacancy V$_{Mo}$ or V$_{S_2}$ marked by a dashed circle is within a 4$\times$4 supercell. (b) Vacancy formation energy E$_{form}$ as a function of biaxial tensile strain.
\label{Fig1} }
\end{figure}
%---------------------------------------------------------------------
\section{results and discussion}
Representative Mo or S$_2$ single vacancy (V$_{Mo}$ or V$_{S_2}$) was created within a 4$\times$4 supercell of MoS$_2$ and marked in dashed circles as shown in Fig. \ref{Fig1}(a). From our calculation, the inter-S distance of 2.80 \AA\ within V$_{Mo}$ is smaller than the 3.19 \AA\ in P-MoS$_2$, while within V$_{S_2}$ the Mo-Mo distance is found to be 3.27 \AA. To compare stability between V$_{Mo}$ and V$_{S_2}$, we calculated vacancy formation energy E$_{form}$ as a function of biaxial tensile strain, as shown in Fig. \ref{Fig1}(b). The E$_{form}$ is defined by N*E$_{atom}$ + E$_d$ - E$_p$, where N and E$_{atom}$ refer to the number and total energy of Mo or S atoms, E$_d$ and E$_p$ are the total energy of defective and P-MoS$_2$, respectively. Both E$_d$ and E$_p$ were calculated under the same level of strain. We found the formation energies without lattice strain are respectively 11.94 eV and 14.26 eV per S$_2$ and Mo vacancy, agreeing with the trend obtained by others\cite{cvd}. V$_{S_2}$ is always obtainable more easily than V$_{Mo}$ within our strain window. Encouragingly, biaxial tensile strain can lower the formation energy, which helps increase the vacancy stability and vacancy creation by particle bombardment in experiment.

\begin{figure}[t]
\includegraphics[width=1.0\columnwidth]{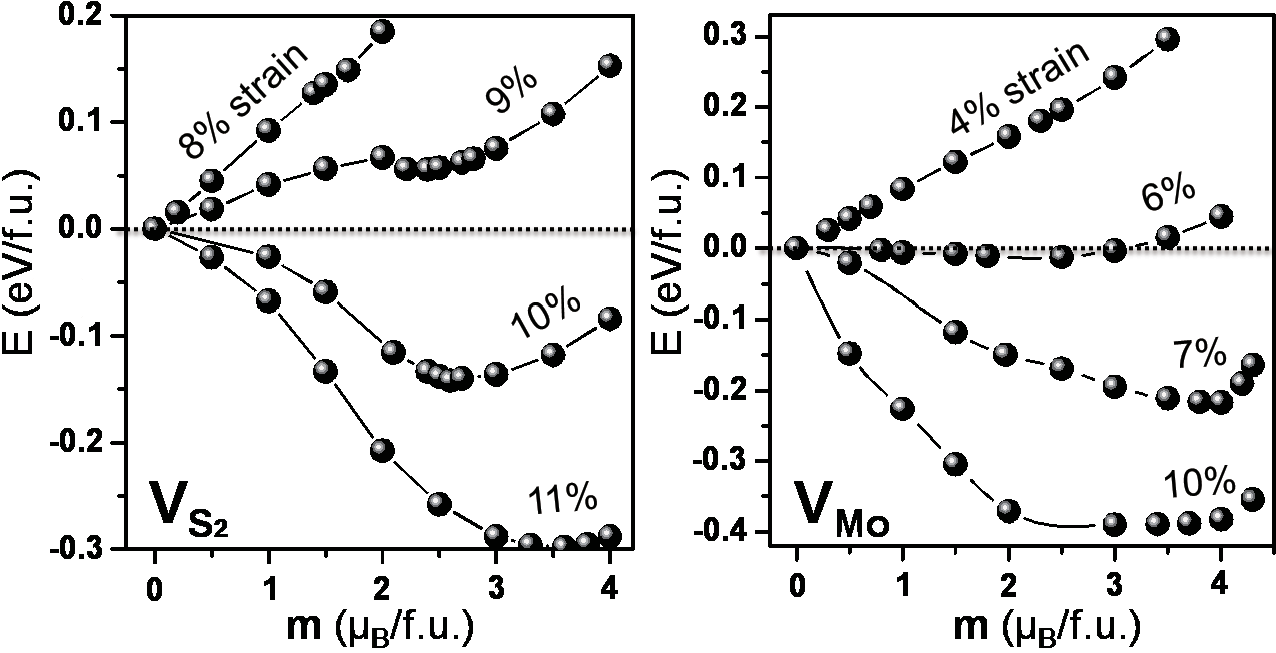}
\caption{The energy gain E-E$_0$ (in eV) as a function of magnetic moment \textbf{m} per formula unit (in $\mu_B$) of (a) V$_{S_2}$- and (b) V$_{Mo}$-MoS$_2$ systems obtained from the fixed spin moment (FSM) method under a series of biaxial tensile strains. E and E$_0$ are the total energies under the same strain with nonzero and zero \textbf{m}, respectively.
\label{Fig2} }
\end{figure}

Besides helping stabilize the vacancy, strain, more importantly, can give rise to magnetism in V$_{Mo}$- and V$_{S_2}$-MoS$_2$ systems. Figure \ref{Fig2} shows energy gain (E-E$_0$) as a function of constrained spin moment \textbf{m} under a series of tensile strains. The energy gain (E-E$_0$) is defined as the total energy difference between nonzero and zero spin moment \textbf{m} states of defected MoS$_2$ under the same strain. Under a small value of strain ($<$10\% for V$_{S_2}$ and $<$7\% for V$_{Mo}$), the energy gain has a global minimum at zero spin moment, showing the V$_{Mo}$- and V$_{S_2}$-MoS$_2$ systems are both nonmagnetic. However, as strain goes beyond some critical value (10\% for V$_{S_2}$, 7\% for V$_{Mo}$)\cite{cristrain}, the energy gain achieves a global minimum at a finite spin moment, suggesting a spontaneous magnetism appearing in both V$_{S_2}$ and V$_{Mo}$ systems. This is in contrast to P-MoS$_2$ which remains nonmagnetic no matter how big the strain is applied. From Fig.\ref{Fig2}(a,b), the magnitude of spin moment is no less than 2 $\mu_B$/vacancy and much bigger than that obtained in defected graphene\cite{rpp73.056501,PhysRevB.75.125408}. It's worth noting that absolute value of energy gain at equilibrium state increases with strain, suggesting a strain-enhanced stability of magnetism.

%--------------------------------------------------------------------
%---------------------------------------------------------------------

\begin{figure}[t]
\includegraphics[width=0.65\columnwidth]{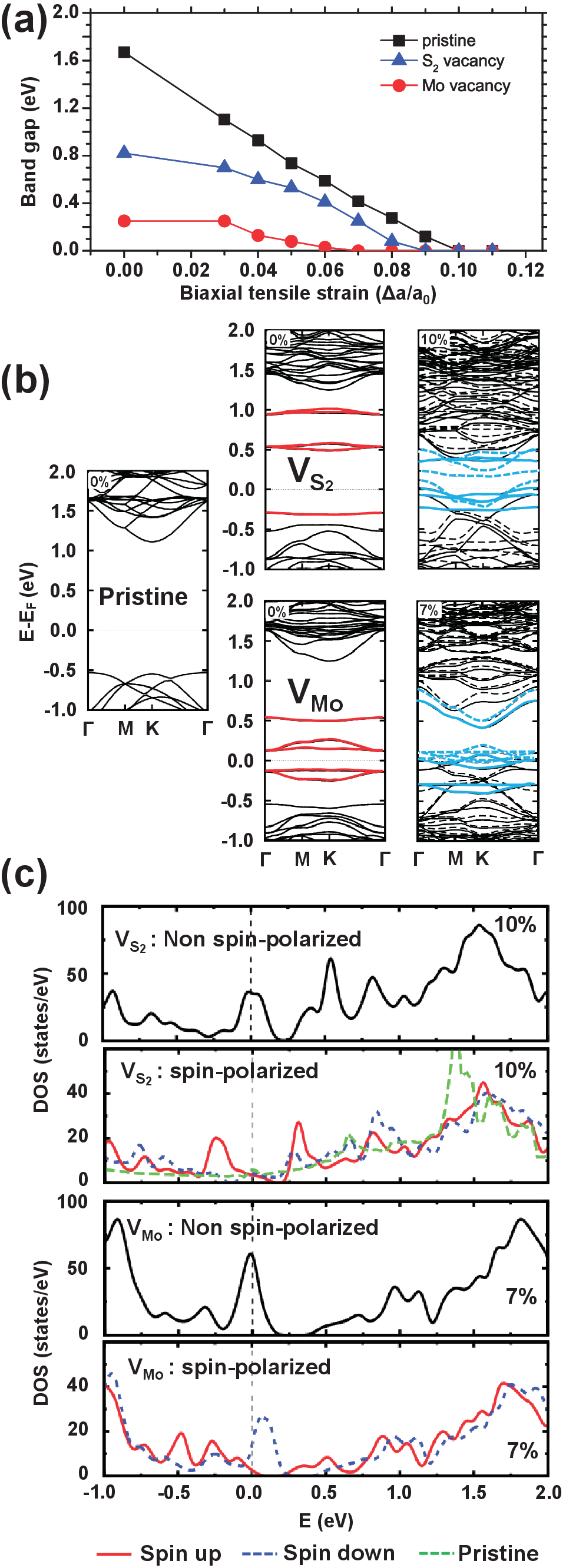}
\caption{(Color online) Electronic structure of V$_{Mo}$, V$_{S_2}$, and pristine MoS$_2$. (a) Band gap as a function of biaxial tensile strain (b) Band structures and (c) densities of states (DOSs) at the corresponding critical strain. Localized defect bands are highlighted in red or blue in (b). The DOS of P-MoS$_2$ marked in green is shifted down for a direct compare with the DOS of defected MoS$_2$ systems, in which spin majority is in solid red and minority in dashed blue line in (c).
\label{Fig3} }
\end{figure}

To get an insight into the strain-induced magnetism, we studied the strain effect on electronic structures. Figure \ref{Fig3} shows the strain dependence of electronic and magnetic properties in the P-, V$_{Mo}$-, and V$_{S_2}$-MoS$_2$ monolayers. All band gaps decrease with increasing biaxial strain, as depicted in Fig. \ref{Fig3}(a). For the P-MoS$_2$, the band gap drops almost linearly with the strain, till it vanishes at 10\%, which is consistent with previous calculations\cite{nanores,hhguojap}. For the V$_{S_2}$- and V$_{Mo}$-MoS$_2$, band gaps were found smaller than that of P-MoS$_2$ and close at 9\% and 7\%, respectively. Interestingly, the critical value of strain for band-gap closing agree quantitatively with that for magnetism emergence in both V$_{S_2}$- and V$_{Mo}$-MoS$_2$, suggesting a close relationship between the magnetism and electron conduction induced by strain.

%--------------------------------------------------------------------
%---------------------------------------------------------------------

To have a microscopic scenario, we first look into the band structures in Fig. \ref{Fig3}(b). Comparing P-MoS$_2$ with either V$_{Mo}$- or V$_{S_2}$-MoS$_2$ under zero strain, we found some nearly dispersionless bands in V$_{Mo}$- or V$_{S_2}$-MoS$_2$ distributing within the band gap of P-MoS$_2$. We considered these almost flat bands as defect levels and highlighted them in red. The defect levels are from the Mo and S atoms in the vicinity of vacancy. Along with increasing strain, they gain a bigger bandwidth. Fig. \ref{Fig3}(b) shows that once these defect levels start to interact with the "bulk" bands around the Fermi level as indicated by blue lines, the spin majority (in solid lines) and minority bands (in dashed lines) split and magnetism appears.

Comparing electronic density of states (DOS) at the critical strain from both non spin-polarized and spin-polarized calculations may give us some idea on how the spin splitting occurs. The DOSs from non spin-polarized calculations are shown by black line in the upper panel of Fig. \ref{Fig3}(c) for each defected system. Fermi level E$_F$ is found to lie in a DOS peak -- a signature of magnetic instability, which therefore gives rise to a spin splitting in the DOS from spin-polarized calculation, as shown in the lower panel of Fig. \ref{Fig3}(c). Besides the peaks close to E$_F$ due to the defect levels, the DOS of both V$_{Mo}$- and V$_{S_2}$-MoS$_2$ remain a DOS background of P-MoS$_2$ under the corresponding strain. For example, as shown in the top panel of Fig. \ref{Fig3}(c), a nearly constant DOS background of V$_{S_2}$-MoS$_2$ spanning an energy range of 1.5 eV was found to agree with the DOS of P-MoS$_2$, displaying an extended electronic state existing near E$_F$ which is obviously due to E$\sim$k$^2$ energy dispersion (qualitatively different from a linear dispersion of Dirac Fermions in graphene\cite{graphene1}). The extended electronic state, coupled with the defect states, results in a possible itinerant origin of magnetism or so-called Stoner ferromagnetism\cite{stoner}. In our case, V$_{Mo}$ has a relatively higher peak value of DOS than that of V$_{S_2}$, and may have a stronger ferromagnetism than V$_{S_2}$, which is actually consistent with the energy stability obtained in Fig.\ref{Fig2}.

%--------------------------------------------------------------------
\begin{figure}[t]
\includegraphics[width=.90\columnwidth]{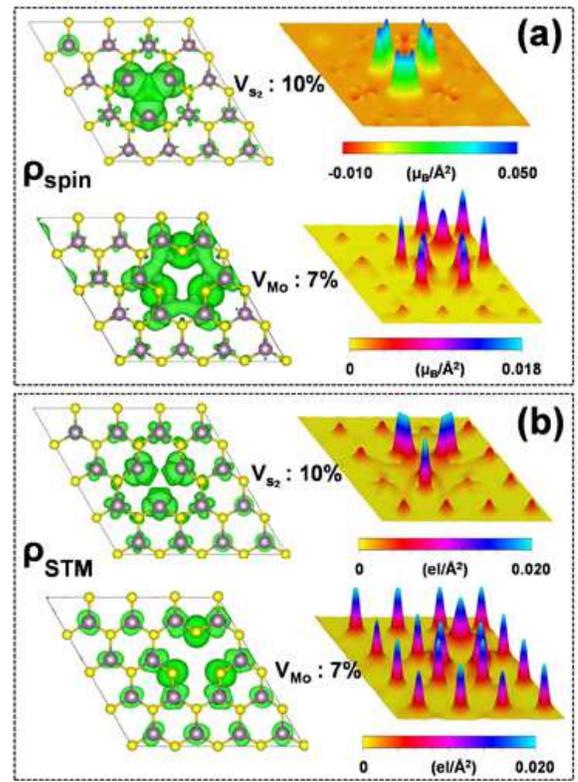}
\caption{(Color online) (a) The spin charge density $\rho_{spin}$ (= $\rho_{\uparrow} - \rho_{\downarrow}$) and (b) total charge density $\rho_{STM}$ (= $\rho_{\uparrow} + \rho_{\downarrow}$) in the energy range from (E$_F$-0.5eV) to E$_F$ at the critical strain where the band gap closes. An extended distribution of $\rho_{spin}$ and $\rho_{STM}$ beyond vacancy region is found, suggesting a possible itinerant magnetism in defected systems.
\label{Fig4} }
\end{figure}
%---------------------------------------------------------------------

To uncover the origin behind the strain-induced magnetism, picture in real space is complementary to (or even more straightforward than) that in reciprocal space. Fig. \ref{Fig4} gives spin charge density $\rho_{spin}$ (= $\rho_{\uparrow}$ - $\rho_{\downarrow}$) and scanning tunneling microscopy (STM) charge density $\rho_{STM}$ (= $\rho_{\uparrow}$ + $\rho_{\downarrow}$) under critical strain for defected MoS$_2$. Both V$_{Mo}$- and V$_{S_2}$-MoS$_2$ monolayer have strong strain-induced localized magnetic moments around the vacancies, as seen from the $\rho_{spin}$ isosurface on the left side and the contour plot on the right side of Fig. \ref{Fig4}(a). The spin magnetization is mainly from the Mo atoms in the vacancy vicinity of V$_{S_2}$-MoS$_2$ and from both Mo and S atoms in the vacancy vicinity of V$_{Mo}$-MoS$_2$. This may give an impression that magnetic moment arises mostly from the dangling bonds of edge atoms. But in the meantime, we found in the contour plot of $\rho_{spin}$ of Fig. \ref{Fig4}(a) that there exists a delocalized distribution of spin charge density over whole unit cell. This relatively small net spin magnetization beyond the vacancy is substantially from all the Mo atomic sites. Actually, the STM charge density integrated from 0.5 eV below E$_F$ to Fermi level in Fig. \ref{Fig4}(b), which contributes to electrical conductivity, also shows a feature of electronic delocalization mainly among Mo sites. Since almost the same amount of electrons are responsible for both the electronic conduction ($\rho_{\uparrow}$+$\rho_{\downarrow}$) and the spin moments($\rho_{\uparrow}$-$\rho_{\downarrow}$), we believe that the strain-induced Stoner magnetism exist in the MoS$_2$ monolayer with atomic vacancies.

%--------------------------------------------------------------------
\begin{figure}[t]
\includegraphics[width=0.8\columnwidth]{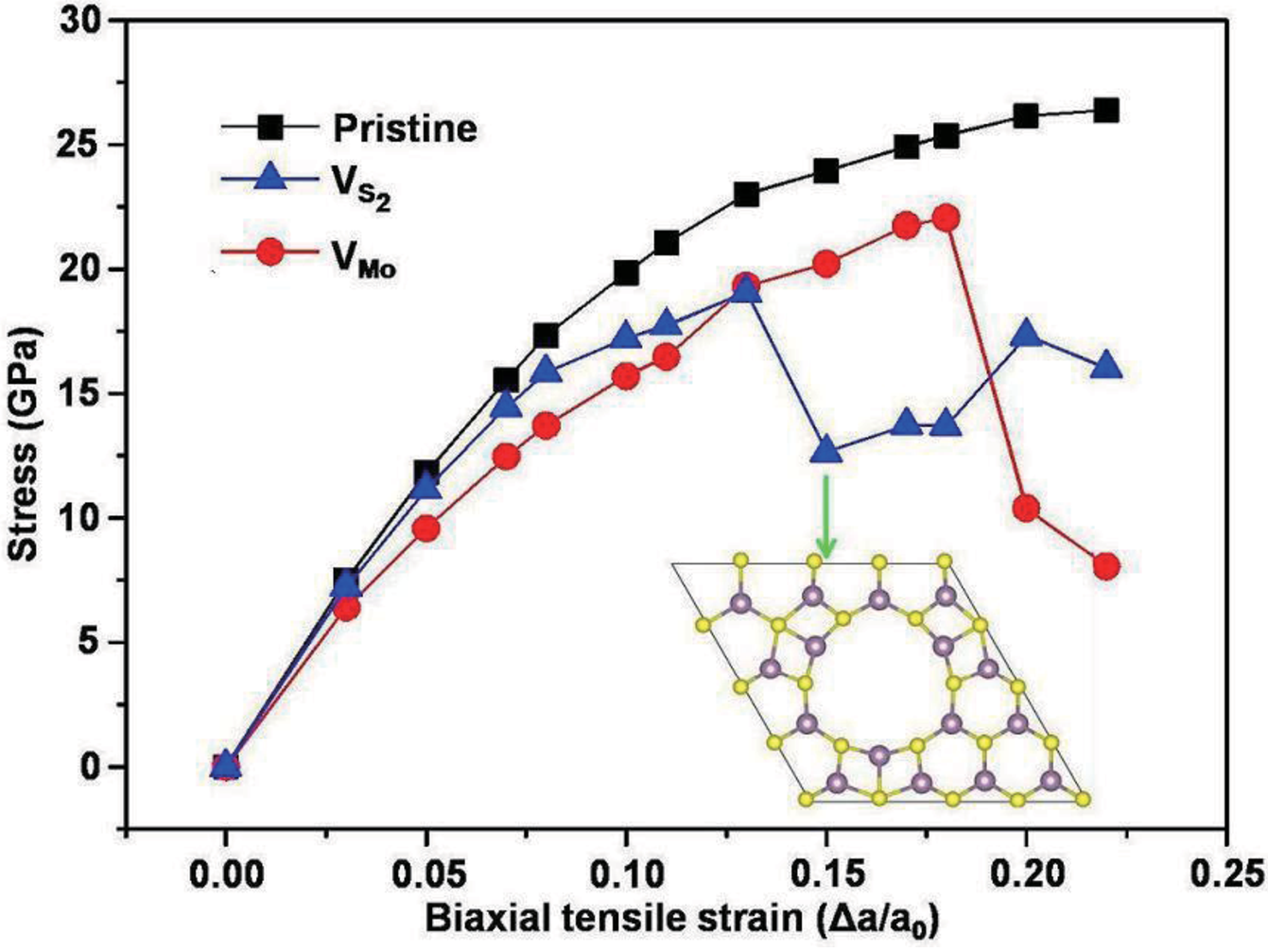}
\caption{The curve of stress vs. biaxial tensile strain for pristine, V$_{Mo}$, and V$_{S_2}$ MoS$_2$ single layer.
\label{Fig5} }
\end{figure}
%---------------------------------------------------------------------

For the response of magnetism to the strain, we found some difference existing between MoS$_2$ monolayer and graphene with single vacancies. Santos et al. \cite{santos} studied graphene with single vacancy and found tensile-strain-enhanced spin moment. In contrast to graphene, V$_{S_2}$- and V$_{Mo}$-MoS$_2$ have no magnetism found till strain induces a semiconductor-metal transition. Magnitude of the spin moment seems not very much dependent on the strain size. We believe that these differences arise mainly from the following aspects. (1) Bigger degree of freedom for atomic relaxation in V$_{S_2}$- and V$_{Mo}$-MoS$_2$ than in graphene. This may help eliminate dangling bonds in the vacancy vicinity of V$_{S_2}$- and V$_{Mo}$-MoS$_2$. Indeed, the bond length of Mo-S in the vicinity of vacancies is about 0.05 \AA\ shorter than the "bulk" one, suggesting a stronger interaction between Mo and S atoms around vacancy than in the "bulk" phase. So no local spin moment is anticipated under zero or small strain. (2) Role of conducting electrons for magnetism. In graphene with single vacancy, besides the delocalized p$_z$ electrons, sp$^2$ dangling bonds\cite{graphenemag,PhysRevFocus.25.6,naturephysics.8.199, santos} were also considered to contribute to the magnetism, even at zero strain. Tensile strain increases the sp$^2$ dangling bonds to enhance the spin moments, whereas compressive strain leads to hybridization between in-plane sp$^2$ and p$_z$ orbitals and therefore to the suppression of magnetism. For MoS$_2$ monolayer, we believe that conducting electrons play a more substantial role in determining magnetism. Tensile strain triggers a semiconductor-metal transition, and more importantly induces hybridization between orbitals in the vicinity of vacancies and delocalized Mo-d$_z$ orbitals\cite{hhguojap}. This hybridization gives rise to magnetic instability in the Fermi level and to itinerant magnetism.

Strain seems indispensable for the magnetism emerging in defected MoS$_2$ monolayer, it is essential to make sure that this happens before structure breaks. We plot a stress vs. biaxial tensile strain curve in Fig. \ref{Fig5}. Fortunately, the critical biaxial strains above which material starts to break ($\sim$18\% and $\sim$14\% for V$_{Mo}$- and V$_{S_2}$-MoS$_2$, respectively, compared to 22\% for P-MoS$_2$ in Fig. \ref{Fig5}) are well beyond the strain for the emergence of magnetism. It is therefore very plausible to test the strain-induced magnetism in MoS$_2$ monolayer with atomic vacancies in experiment.

\section{conclusion}
In conclusion, we studied the magnetic, electronic and mechanical properties of single-layer MoS$_2$ with S$_2$ and Mo vacancy under biaxial tensile strain, based on \emph{ab. initio.} density functional calculations. A spontaneous magnetization appears as band gap closes under certain biaxial tensile strain. We understand the strain-induced magnetism for V$_{Mo}$- and V$_{S_2}$-MoS$_2$ from the itinerant origin based on the analysis of electronic structures, density of states and charge density at the critical strain. The delocalized electrons of Mo are believed to be responsible for this unique itinerant magnetism in the defected MoS$_2$. We also discussed the possibility to experimentally test our theoretical predictions. The strain-induced magnetic moments may provide MoS$_2$ an opportunity to the design of magnetic-switching or logic devices.

\begin{acknowledgments}
The work is supported by the National Basic Research Program (No. 2012CB933103) of China, Ministry of Science and Technology China and the NSFC under Grant No. 11004201, 51331006. T.Y. acknowledges the IMR SYNL-Young Merit Scholars for support.
\end{acknowledgments}

%\bibliography{d-MoS2}

\begin{thebibliography}{34}
\bibitem [{cri()}]{cristrain}%
  \BibitemOpen
  \href@noop {} {}\bibinfo {note} {Here, V$_{S_2}$ at 9\% or V$_{Mo}$ at 6\% is
  not treated to be magnetic since the global minimum state occurs at zero spin
  moment.}\BibitemShut {Stop}%
\end{thebibliography}

\end{document}